# Development and validation of an optimal GATE model for proton pencil-beam scanning delivery

A. Asadi[1], A. Akhavanallaf[2], S. A. Hosseini[1], N. vosoughi[1], H. Zaidi[2,3,4,5]

[1]Department of Energy Engineering, Sharif University of Technology, Tehran, Iran, Zip code: 8639-11365

[2]Division of Nuclear Medicine and Molecular Imaging, Geneva University Hospital, CH-1211 Geneva 4, Switzerland

[3]Geneva University Neurocenter, Geneva University, CH-1205 Geneva, Switzerland

[4]Department of Nuclear Medicine and Molecular Imaging, University of Groningen, University Medical Center Groningen, 9700 RB Groningen, Netherlands

[5]Department of Nuclear Medicine, University of Southern Denmark, DK-500, Odense, Denmark

**Short running title:** pencil beam proton therapy Monte Carlo simulator

**Abstract**

**Objective:** To develop and validate an independent Monet Carlo (MC)-based dose calculation engine to support for software verification of treatment planning systems (TPS) and quality assurance (QA) workflow.

**Methode:** GATE Monte Carlo toolkit was employed to simulate a fixed horizontal active scan-based proton beam delivery (SIEMENS IONTRIS). Within the nozzle, two primary and secondary dose monitors have been designed allowing to compare the accuracy of dose estimation from MC simulation with respect to physical quality assurance measurements. The developed beam model was validated against a series of commissioning measurements using pinpoint chambers and 2D array ionization chambers in terms of lateral profiles and depth dose distributions. Furthermore, beam delivery module and treatment planning has been validated against the literature deploying various clinical test cases of AAPM TG- 119 (c-shape phantom) and a prostate patirnt.

**Result:** MC simulation showed an excellent agreement with measurements in the lateral depth-dose parameters and spread-out Bragg peak (SOBP) characteristics within maximum relative error of 0.95% in range, 3.4% in entrance to peak ratio, 2.3 % in mean point to point, and 0.852% in peak location. Mean relative absolute difference between MC simulation and the measurement in terms of absorbed dose in SOBP region was 0.93%±0.88%. Clinical phantom study showed a good agreement compared to a commercial treatment planning system (relative error for TG-119 PTV-$D_{95}$ ~ 1.8%; and for prostate PTV-$D_{95}$ ~ -0.6%).

**Conclusion:** The results confirm the capability of GATE simulation as a reliable surrogate for verifying TPS dose maps prior to patient treatment.

# 1. Introduction

There is a globally growing interest in using particles compared to conventional radiation therapy owing to the inherent potentials of particles in dose-painting [1] through escalating the delivered dose to target while sparing normal tissues. Recently, active scanning proton therapy called Pencil Beam Scanning (PBS) has become a reliable and preferred method of cancer treatment compared to conventional passive scattering proton therapy technique, mainly because of providing an elegant conformal dose distribution and facilitated beam delivery without need for multiple field-specific scatterer mechanical hardware [2, 3]. Commercially available proton therapy facilities, mostly employ semi-analytic algorithms for dose planning. Treatment planning system (TPS) simplifications in dose planning in complex situations such as utilization of aperture (reducing the lateral penumbra) or range shifter (shallow tumor treatments), may lead to large errors between prescribed dose and delivered dose [4]. To ensure the correct delivery of the planned dose through TPS, pre-treatment dose verifications called patient-specific quality assurance is recommended for clinical workflows  ADDIN EN.CITE [5, 6]. However, due to the growing demand for proton therapy and increasing the treatment time, such procedures are less feasible in routine clinical practice. In this context, independent validated Monte Carlo simulation tools can be properly utilized for patient-specific dose monitoring prior to the treatment. Multiple Monte Carlo codes such as FLUKA [7, 8], MCNP [9], Geant4 [10], and TOPAS [11] have been employed for proton beam delivery simulation allowing users to validate and examine various aspects of the beam-material interactions  ADDIN EN.CITE [12-16]. A substantial body of literature has been reported various MC models for both passive scattering and active scanning systems of cyclotron and synchrotron-based proton therapy facilities. Almhagen et al [17]. simulated an active proton scanning design to investigate the potential of MC-based models in software quality assurance and patient-motion studies. Grevillot et al [18], developed an MC-based simulator for modeling the Ion Beam Application (IBA) active scanning proton therapy system using Beam Data Library. They reported a slight difference between the simulation results with respect to the measurement using 2D-array ionization chambers.

In this work, we developed a unified framework for patient-specific quality assurance calculation based on GATE Monte Carlo toolkit. A dosimetric comparison between the simulation results and the measurement using pinpoint chambers and 2D-array ionization chambers has been conducted.

The developed simulator has been further validated against a commercial TPS through a clinical study using TG-119 c-shape phantom, and prostate patient.

## 2. Materials and Methods

### 2.1 system description

Shanghai Advanced Proton Therapy (SAPT) facility, a synchrotron-based active scanning proton therapy system, was simulated. Fig 1. depicts the geometrical characteristics of SIEMENS IONTRIS system at SAPT. In this system proton beams are extracted from synchrotron and drifted to the nozzle, by using the paired scanning magnets in horizontal (X) and vertical (Y) directions. Proton beam spot is moved around the isocenter with energies between 70-235 MeV [19]. The advantage of this technique against scattering-based technique is that a range shifter is not required to shape the beam to the tumor volume, because the synchrotron accelerate the protons slowly and conform the tumor dose in the dimension lateral to the beam [19]. The characteristics of this system are listed in Table 1.

**Table 1.** The characteristics of SAPT proton therapy system [19].

| Item | value |
|---|---|
| Energy (MeV) | 70.0-235.0 |
| Field size (cm$^2$) | 30.0×40.0 |
| Scanning magnet x to isocenter distance (cm) | 287.0 |
| Scanning magnet x to isocenter distance (cm) | 247.0 |
| Nozzle to isocenter distance (cm) | 40.0 |
| Scan speed in x (cm/ms) | 2.0 |
| Scan speed in y (cm/ms) | 0.5 |
| Dose rate (Gy/min) | 2.0 |

Dose delivered to the phantom is monitored in real time by using two parallel-plate ionization chambers. Spot size and beam optic are measured by using the position detectors. Unlike discrete scan mode (pixel scan), IONTRIS provided a continuous beam scan mode (raster scan).

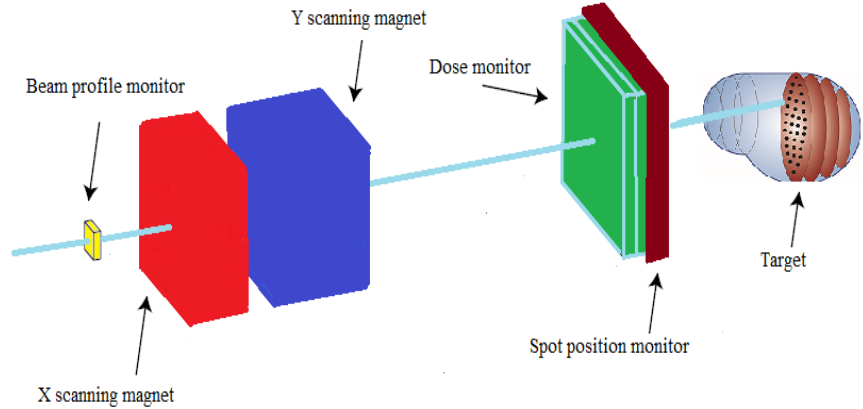

**Fig. 1**. The simple schematic of SAPT nozzle[19].

## 2.2 GATE simulation

In this study, GATE toolkit was employed since it has been previously validated for clinical operation of particle therapy [18]. Proton beam and monitoring devices has been simulated based on the method proposed by Grevilot et al [18]. The initial beam (source plane) was set at the nozzle entrance immediately before the vacuum window, and angular spread particle sampling strategy was adopted. The geometry of ion chamber was simplified to be water with the corresponding water equivalent depth reported by the manufacturer. The multiwire peak chamber was described as water covered by a 3-micrometer tungsten layer that represents scattering of low energy protons in the high-Z tungsten wires. The cut value on the particle transport was set at 0.1 mm. Fig 2. Schematically illustrates the simulation workflow. In this study, case-specific dose modeling has been conducted in three main parts: (a) feeding phantom (water or patient anatomical image) geometry into the simulator using the calibration curve of CT-to-stopping power; (b) treatment plan optimization using the irradiation criteria (prescribed dose to target and restrictions) and contours obtained from DICOMRT-structure; and (c) recording 3D dose map and extracting clinically relevant dose-volume parameters.

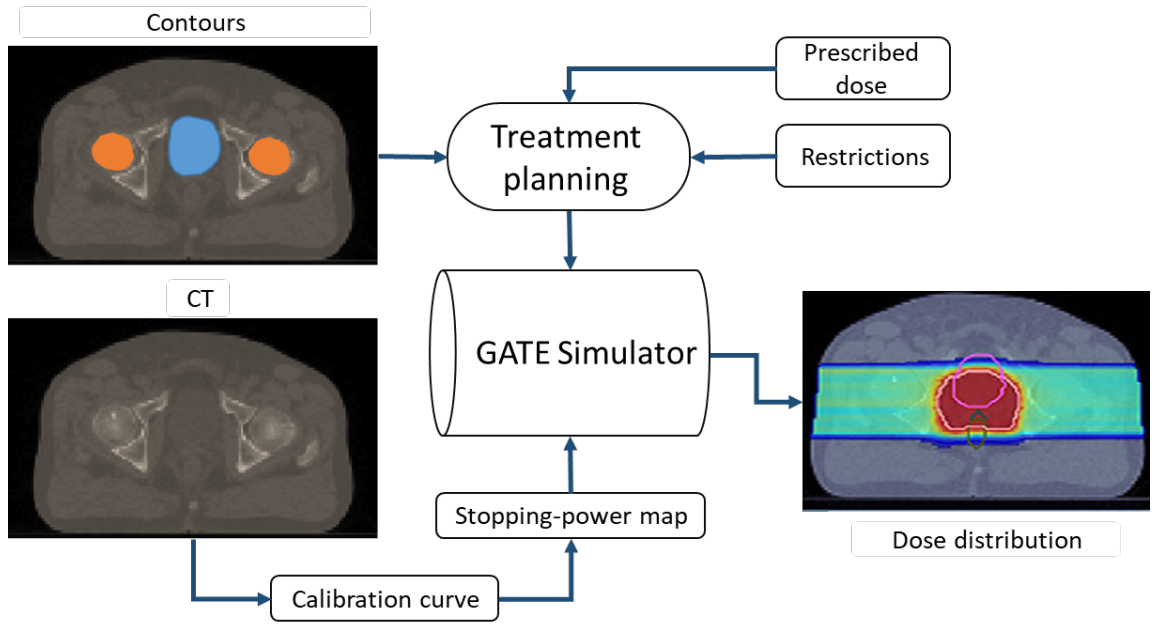

**Fig. 2.** Schematic simulation workflow.

### 2.3 Beam characterization and simulation validation

The developed MC simulator has been benchmarked against experimental measurements, reported by Hang Shu, et al [19].

### 2.3.1 Integral Depth Dose (IDD)

Dose distribution curves for six energy ranges between 130.1-235 MeV are obtained by water phantom (MP30PL) and Bragg peak chamber (Model 34070,PTW) with nominal sensitive chamber volume of 10.5 cm in diameter and water equivalent thickness of 0.4 cm. The PEAKFINDER device was positioned 5 cm downstream from the nozzle exit with a radius of 4.08 cm. The energy of the simulated proton beams was adjusted to match the range (defined as R80) of measurement for each IDD. Furthermore, the parameter of clinical range (R90) was calculated [18].

### 2.3.2 Spot size

The spot size was determined using both a radiographic film (Gafchromic EBT Film) and commercial 2D scintillator (lynx). The EBT film and Lynx were exposed in the air at four locations: the isocenter plane, 40 cm downstream isocenter, isocenter, and 20 cm upstream the isocenter plane. The Gafchromic film was simulated through a very thin titanium plate that located perpendicular to the beam path. A phase space actor was used to record the input particles to the volume to reconstruct the spot size.

### 2.3.3 Full Widths at Half Maximum (FWHM)

FWHM in X and Y directions were derived based on the lateral profile to quantify the beam spot. A total of 15 energy spot sizes were calculated. The proton beam was defined as a point source, while a Gaussian-shaped beam angular spread in X and Y directions was implemented in the simulation. 2D dose distributions were scored at the same locations. The resolution of scoring along the central beam axis was set to 0.36 mm to match the spacing of measurements. The angular spread in the X and Y directions was adjusted to match the measured FWHM, and two fourth-order polynomials were used to fit the angular spread FHWM in X and Y directions as a function of the nominal energy.

### 2.3.4 1-D dose profiles in water

Range modulated plans were generated by the treatment planning system (V13B, Syngo, Siemens) in a water tank. The target size was set to 3×3×3 cm$^3$ and 6×6×6 cm$^3$ at the center of the cube of 5 cm edge and 20 cm edge, respectively. The prescribed dose was set to 0.5 Gy. The range shifter was located 20 cm before the surface of the water tank for the range modulated plan with a shallow target. The plan was delivered to the water tank (MP3-P, PTW-Freiburg). Measurements were performed by using 24 PinPoint chambers (T31015, PTW-Freiburg) at different positions to obtain sufficient resolution for depth and lateral dose profiles. Beam delivery parameters including beam energy, scanning cosine angle for X and Y directions, particle number of each scanning spot, and the energy selection plan for modulation region were imported to simulation code. The inverse planning optimization algorithm proposed by Bourhaleb et al. [20] was implemented to achieve a uniform dose to the target volume. The absorbed dose distribution per particle was scored. The step grid for scoring was set to 2 mm in X and Y axis, whereas it was set to 1 mm for Z (central) axis. The GATE user routine was used to normalize the GATE results to the absorbed dose.

### 2.4 Clinical phantom study

We further validated our MC-based simulator against a commercial TPS. Clinical phantom and case studies reported by Daniel et al [21] have been modeled and compared against the published Reference. AAPM (TG-119) C-shape phantom was imported to the GATE TPS source, where the C-shape PTV intricate around a core structure whose outer surface is 0.5 cm from the inner surface of the PTV. In addition, a prostate case-study, taken from Daniel et al [21] work has been simulated. The Structure set contains the Target (PTV), bladder, rectum, femur, and body. Accordingly, The TG-119 phantom was prescribed with 50 GyRBE dose to the PTV (target) and the maximum dose to the 5 % of organ at risk with 10 GyRBE [21]. A single proton field was set on the target to maximize the biological effect. A constant factor of 1.1 was applied to the physical dose to hypothesize the relative biological effect of protons. For optimization of spot and beam selection, the inverse planning optimization algorithm [20] was used to uniformly cover the target. The prostate phantom was prescribed with 78 GyRBE dose to the PTV region using 2 parallel opposed proton fields. Table 2 summarizes the radiation treatment objectives and restrictions.

**Table 2.** Treatment plan objectives for TG-119 [22] and prostate [23] cases.

|          | Structure | Parameter | Goal (%) |
|----------|-----------|-----------|----------|
| C-shape  | PTV       | V10       | <55      |
|          |           | V99       | 50       |
|          | core      | V5        | 10       |
| Prostate | bladder   | V70       | <35      |
|          |           | V50       | <60      |
|          | rectum    | V70       | <30      |
|          |           | V50       | <50      |
|          | femur     | V50       | <5       |

Quantitative comparison of dose-volume histogram parameters was calculated based on the relative differences as following

$$RPE= \qquad (1)$$

where, $D_i$ and $D_i'$ represents the dose parameter for pencil beam scanning system and those obtained from Daniel et al [21], respectively.

## 3. Results

### 3.1 IDDs

The difference of beam range (R80) between the experimental data and the simulation data in all 6 simulated energies was less than 1%. The energy dissipation was of the order of 0.2% , which agrees with previous results reported by Grevillot et al [18]. In this work, a set of depth-dose curves was calculated for 6 energies. An example of the obtained curves for 6 energy ranges (130.1, 161.1, 179.9, 202, 219.2, and 235 MeV) is shown in Fig. 3. The peak to entrance dose ratio, mean point-to-point, Bragg-peak location, and range deviation between the measured data and fitted simulation were compared (Fig. 4). The maximum deviation of the beam range between the measured and simulated data was 0.85% (at the energy of, 235.0 MeV). The mean range deviation was also 0.468 %. For mean point-to-point dose difference between the measurements and fitted simulation, the mean value and the maximum value were 0.53% and 0.72%, respectively. For all cases, the mean point-to-point deviation value was lower than 1%. The maximum peak-to-entrance dose deviation between the measurement and fitted simulation was 3.5 % for proton beam energy 202 MeV.

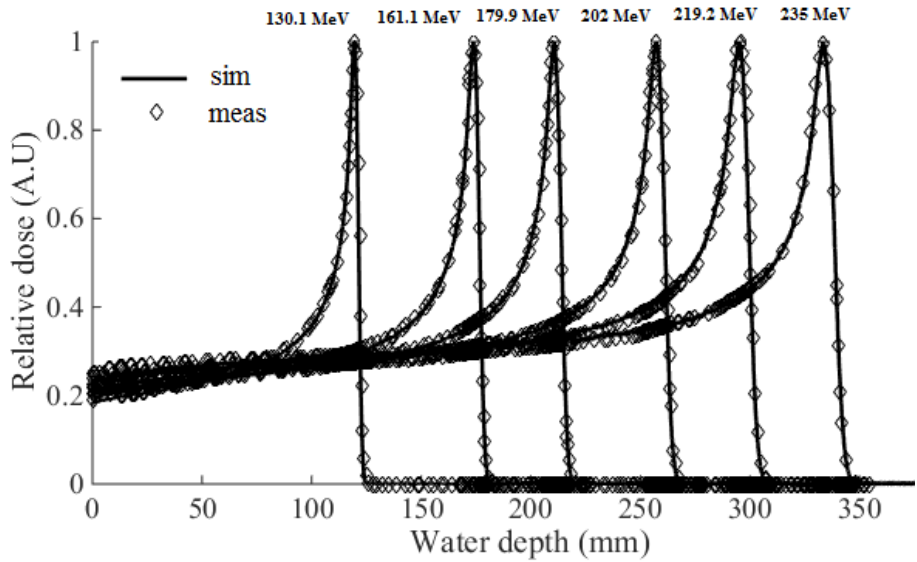

**Fig. 3.** Simulated and measured IDD′s at different energies.

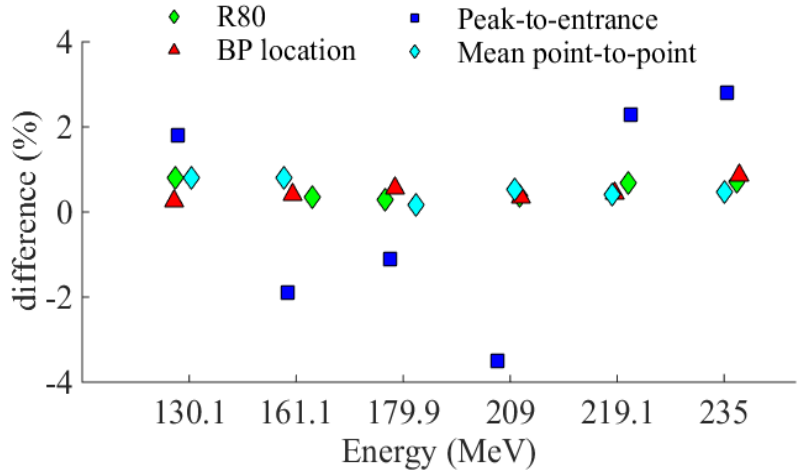

**Fig .4.** Comparison of range, Entrance-to-peak dose ratio, mean point-to-point, and Bragg-peak location deviation between the measurements and simulation.

## 3.2 In air spot size

The deviation between the simulated and measured spot sizes in air, at the isocenter was within ±0.8 mm in the X and Y directions for all 6 proton energies. We recalculated the results based on a fitted angular distribution. The deviation between the results of measurements and fitted simulation at the isocenter is shown in Fig. 5. The simulated and measured spot sizes in X direction at various distance from the isocenter for two energies (121.08 MeV and 221.07 MeV) were presented in Fig. 4.b. The mean difference was 0.08 mm at the isocenter, and the maximum difference was 0.6503 mm (at the proton energy of 235.0 MeV). The spot size were reproduced within ±0.8 mm for the two positions around the isocenter. The mean difference were -4.52% at the 40 cm upstream of the isocenter and 2.48% at the 20 cm downstream of the isocenter.

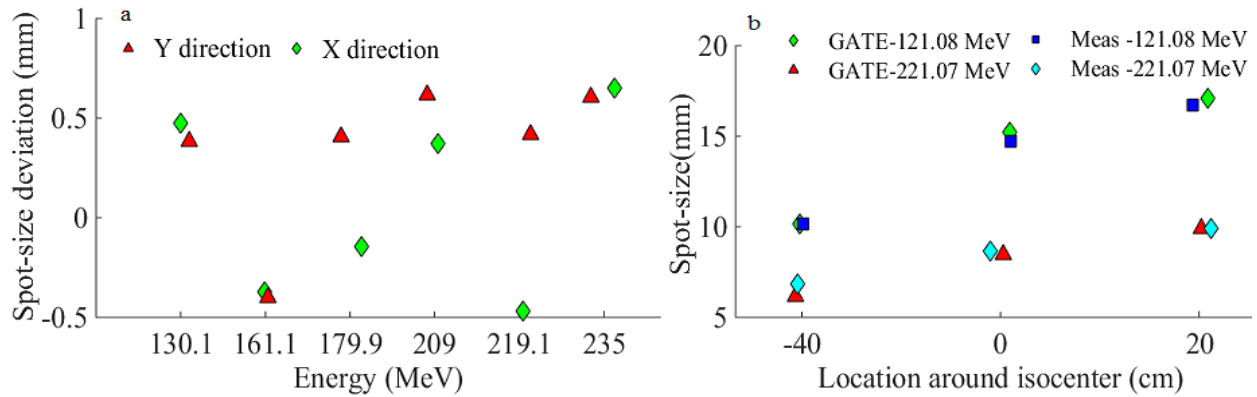

**Fig 5.** (a) Differences between simulated and measured spot size at isocenter for both X and Y directions for different energies. (b) Sample of the simulated and measured spot sizes in X direction for energies of 121.08 and 221.07 MeV at different locations around isocenter.

## 3.3 Comparison of 1-D dose profiles

Measured and MC-simulated depth-dose curve along the central axis and transverse beam profile at the center of the SOBP cubes are shown in Fig. 6. The results show good agreement between the simulation and measurements. The clinical range difference was 0.22 mm, and the 80%-20% distal fall off value difference was 0.11 mm. The mean (SD) and maximum dose differences of the depth dose profile were 0.93% (0.88%) and 3.53%, respectively, for the measured points except for points at the distal edge of the SOBP. The dose differences of the lateral dose profile between the simulation and measurements at the FWHM dose level were within 2.30%.

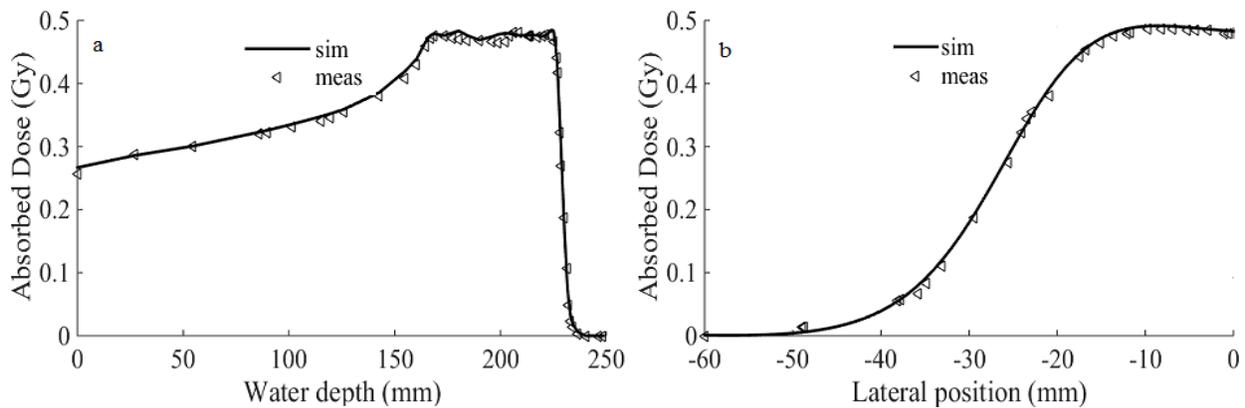

**Fig 6.** Comparison of the SOBP plan in (a) Longitudinal profile and (b) transverse profile at the center of the water phantom.

## 3.4 Clinical phantom evaluation

Fig. 7 Shows the dose distribution for TG-119 phantom and prostate case for the pencil beam scanning proton therapy plan along with DVH analysis for target, and OARs. The color wash is normalized (TG-119: 30 fraction, each 1.66 Gy, and prostate case: 78 Gy in 39 fraction). Table 3 (a and b), represent the DVH-driven parameters obtained from simulation compared to [21].

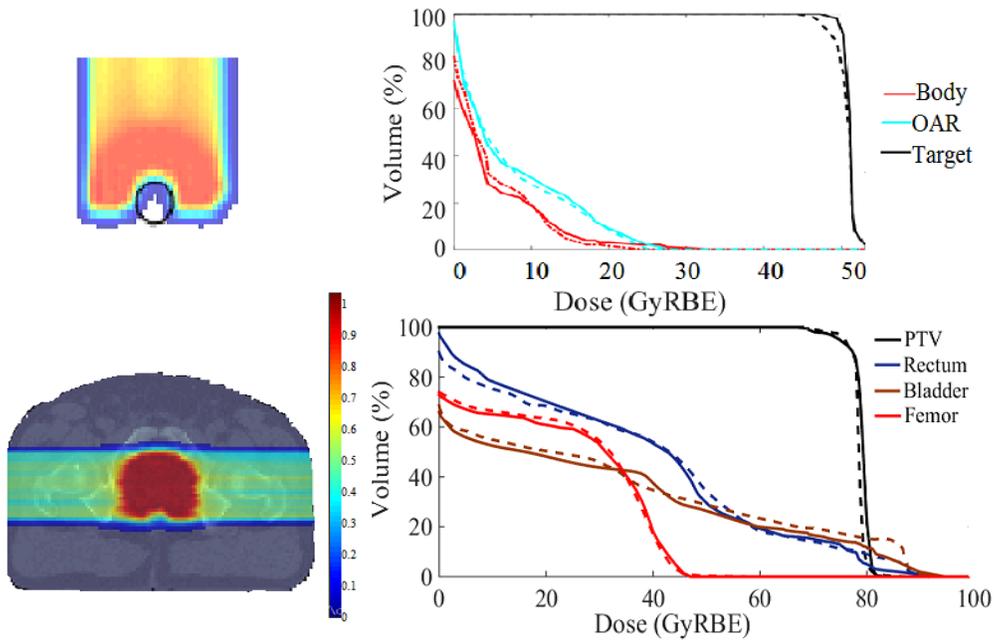

**Fig 7.** Simulated **d**ose distribution for TG-119 phantom (top) and prostate (bottom) cases in the pencil beam scanning proton therapy plan along with DVH plots for target and OARs compared to [21].

| ROI | $D_5$ (Gy) | | | $D_{95}$ (Gy) | | |
|---|---|---|---|---|---|---|
| | simulation | Daniel et al | Dif (%) | simulation | Daniel et al | Dif (%) |
| Target | 52.25 | 49.4 | 5.4 | 50.3 | 49.4 | 1.8 |
| OAR | 12.3 | 11.3 | 8 | 0 | 0 | 0 |

**Table 3.a**. Quantitative analysis for DVH parameters in TG-119 phantom

**Table 3.b** Quantitative analysis for DVH parameters in prostate case

| Structure | | Goal [23] | Daniel et al | sim | Dif (%) |
|---|---|---|---|---|---|
| **Target** | $D_{max}$ (Gy) | -- | 82.5 | 83.0 | -0.60 |
| | $D_{95}$ (Gy) | -- | 75.5 | 75.0 | 0.66 |
| **Bladder** | $D_{max}$ (Gy) | -- | 91.9 | 88.5 | 3.6 |
| | $V_{70}$ (Gy) | <35 | 19.36 | 19.5 | 0.72 |
| **Rectum** | $D_{max}$ (Gy) | -- | 90.1 | 90.0 | 0.10 |
| | $V_{70}$ (%) | <25 | 11.6 | 11.9 | -2.58 |
| **Femur** | $V_{50}$ (%) | <5 | 0 | 0 | 0 |

## 4. Discussion

SAPT system has been simulated using GATE code. The accuracy of the simulation has been investigated through multiple parameters, i.e. dose-depth curves, spread out Bragg-peak, range, peak location, mean point to point difference, and spot size in the air. The results show that there is a good agreement between the experimental data and results of the simulation. The results of the domain spread plan show that the deviation from the experimental measurement is about 1.3%. The modeling was performed based on the measurement of the transverse profile of the dose in the water phantom with dimensions of $(40 \text{ cm})^3$. The deviation between the measured and simulated range was less than 1%. The maximum difference between the dose at the peak and the input dose was -3.5%. It can be originated from the physical parameters defined in the simulation (physics-list). In addition, there are different reports for ionization potential that directly influence on the simulation results [24]. According to ICRU, a potential of 78 ev has been implemented in this simulation  ADDIN EN.CITE [25-27]. The mean difference of spot size were up to -4.52% at 40 cm upstream of the isocenter. According to [18], these differences are within quality assurance limits and considered acceptable. We further validate our work against a commercial TPS, using clinical phantom/case studies. The results show that the modeling performed using the GATE simulator combined with treatment planning algorithm agrees well with the experimental data. Hence the use of the Monte Carlo GATE code can be considered as a gold standard in dosimetric calculations for active scanning proton therapy.

## 5. Conclusion

The developed MC-based simulator has been developed for dose calculation in Pencil Beam scanning proton therapy. Simulated plans for water phantoms and clinical cases, agree well with experimental measurements (obtained froman ionization chamber and 2D array) and the commercial TPS. Results demonstrate the capability of the developed simulator as an independent software QA program that can be implemented in the clinical workflow.

## References

 ADDIN EN.REFLIST [1]   A. Lomax, "Intensity modulation methods for proton radiotherapy," *Physics in Medicine & Biology,* vol. 44, no. 1, p. 185, 1999.
[2]	J. S. Loeffler and M. Durante, "Charged particle therapy―optimization, challenges and future directions," *Nature reviews Clinical oncology,* vol. 10, no. 7, pp. 411-424, 2013.
[3]	E. Pedroni *et al.*, "Initial experience of using an active beam delivery technique at PSI," *Strahlentherapie und Onkologie,* vol. 175, no. 2, pp. 18-20, 1999.

[4]   D. Nichiporov, V. Moskvin, L. Fanelli, and I. Das, "Range shift and dose perturbation with high-density materials in proton beam therapy," *Nuclear Instruments and Methods in Physics Research Section B: Beam Interactions with Materials and Atoms,* vol. 269, no. 22, pp. 2685-2692, 2011.

[5]   B. Arjomandy, N. Sahoo, G. Ciangaru, R. Zhu, X. Song, and M. Gillin, "Verification of patient-specific dose distributions in proton therapy using a commercial two-dimensional ion chamber array," *Medical physics,* vol. 37, no. 11, pp. 5831-5837, 2010.

[6]   X. R. Zhu *et al.*, "Patient-specific quality assurance for prostate cancer patients receiving spot scanning proton therapy using single-field uniform dose," *International Journal of Radiation Oncology\* Biology\* Physics,* vol. 81, no. 2, pp. 552-559, 2011.

[7]   G. Battistoni *et al.*, "The FLUKA code: an accurate simulation tool for particle therapy," *Frontiers in oncology,* vol. 6, p. 116, 2016.

[8]   A. Ferrari, J. Ranft, P. R. Sala, and A. Fassò, *FLUKA: A multi-particle transport code (Program version 2005)* (no. CERN-2005-10). Cern, 2005.

[9]   J. M. Ryckman, "Using MCNPX to calculate primary and secondary dose in proton therapy," Georgia Institute of Technology, 2011.

[10]  J. Allison *et al.*, "Geant4 developments and applications," *IEEE Transactions on nuclear science,* vol. 53, no. 1, pp. 270-278, 2006.

[11]  J. Perl, J. Shin, J. Schümann, B. Faddegon, and H. Paganetti, "TOPAS: an innovative proton Monte Carlo platform for research and clinical applications," *Medical physics,* vol. 39, no. 11, pp. 6818-6837, 2012.

[12]  W. Newhauser *et al.*, "Monte Carlo simulations for configuring and testing an analytical proton dose-calculation algorithm," *Physics in Medicine & Biology,* vol. 52, no. 15, p. 4569, 2007.

[13]  H. Paganetti, "Monte Carlo method to study the proton fluence for treatment planning," *Medical physics,* vol. 25, no. 12, pp. 2370-2375, 1998.

[14]  H. Paganetti, "Monte Carlo calculations for absolute dosimetry to determine machine outputs for proton therapy fields," *Physics in Medicine & Biology,* vol. 51, no. 11, p. 2801, 2006.

[15]  H. Paganetti, H. Jiang, S. Y. Lee, and H. Kooy, "Accurate Monte Carlo simulations for nozzle design, commissioning and quality assurance for a proton radiation therapy facility," *Medical physics,* vol. 31, no. 7, pp. 2107-2118, 2004.

[16]  K. Parodi *et al.*, "Monte Carlo simulations to support start-up and treatment planning of scanned proton and carbon ion therapy at a synchrotron-based facility," *Physics in Medicine & Biology,* vol. 57, no. 12, p. 3759, 2012.

[17]  E. Almhagen, D. J. Boersma, H. Nyström, and A. Ahnesjö, "A beam model for focused proton pencil beams," *Physica Medica,* vol. 52, pp. 27-32, 2018.

[18]  L. Grevillot, D. Bertrand, F. Dessy, N. Freud, and D. Sarrut, "A Monte Carlo pencil beam scanning model for proton treatment plan simulation using GATE/GEANT4," *Physics in Medicine & Biology,* vol. 56, no. 16, p. 5203, 2011.

[19]  H. Shu *et al.*, "Scanned Proton Beam Performance and Calibration of the Shanghai Advanced Proton Therapy Facility," *MethodsX,* vol. 6, pp. 1933-1943, 2019.

[20]  F. Bourhaleb *et al.*, "A treatment planning code for inverse planning and 3D optimization in hadrontherapy," *Computers in biology and medicine,* vol. 38, no. 9, pp. 990-999, 2008.

[21]  D. Sánchez-Parcerisa, M. López-Aguirre, A. Dolcet Llerena, and J. M. Udías, "MultiRBE: Treatment planning for protons with selective radiobiological effectiveness," *Medical physics,* vol. 46, no. 9, pp. 4276-4284, 2019.

[22]  D. K. Mynampati, R. Yaparpalvi, L. Hong, H. C. Kuo, and D. Mah, "Application of AAPM TG 119 to volumetric arc therapy (VMAT)," *Journal of applied clinical medical physics,* vol. 13, no. 5, pp. 108-116, 2012.

[23]  N. P. Mendenhall *et al.*, "Early outcomes from three prospective trials of image-guided proton therapy for prostate cancer," *International Journal of Radiation Oncology\* Biology\* Physics,* vol. 82, no. 1, pp. 213-221, 2012.

[24] P. Andreo, "On the clinical spatial resolution achievable with protons and heavier charged particle radiotherapy beams," *Physics in Medicine & Biology,* vol. 54, no. 11, p. N205, 2009.

[25] D. K. Brice, "Book Review: Stopping powers for electrons and positrons (ICRU report 37; International commission on radiation units and measurements, Bethesda, Maryland, USA, 1984). pp. viii+ 267, $24.00; ISBN 0-913394-31-9," *Nuclear Instruments and Methods in Physics Research B,* vol. 12, no. 1, pp. 187-188, 1985.

[26] B. Siebert and H. Schuhmacher, "Quality factors, ambient and personal dose equivalent for neutrons, based on the new ICRU stopping power data for protons and alpha particles," *Radiation Protection Dosimetry,* vol. 58, no. 3, pp. 177-183, 1995.

[27] D. Schardt, P. Steidl, M. Krämer, U. Weber, K. Parodi, and S. Brons, "Precision Bragg-curve measurements for light-ion beams in water," *GSI Scientific Report,* vol. 373, 2007.